\documentstyle[12pt]{article}   

\large
\newcommand{\beq}{\begin{equation}}
\newcommand{\eeq}{\end{equation}}

\makeatletter
\@addtoreset{equation}{section}
\makeatother

\def\be{\begin{equation}}
\def\ee{\end{equation}}
\def\ba{\begin{eqnarray}}
\def\ea{\end{eqnarray}}

\def\agb{{\overline {{\cal A}/{\cal G}}}}

\def\Comp{{\mathchoice
{\setbox0=\hbox{$\displaystyle\rm C$}\hbox{\hbox to0pt
{\kern0.4\wd0\vrule height0.9\ht0\hss}\box0}}
{\setbox0=\hbox{$\textstyle\rm C$}\hbox{\hbox to0pt
{\kern0.4\wd0\vrule height0.9\ht0\hss}\box0}}
{\setbox0=\hbox{$\scriptstyle\rm C$}\hbox{\hbox to0pt
{\kern0.4\wd0\vrule height0.9\ht0\hss}\box0}}
{\setbox0=\hbox{$\scriptscriptstyle\rm C$}\hbox{\hbox to0pt
{\kern0.4\wd0\vrule height0.9\ht0\hss}\box0}}}}
\def\Co{{\mathchoice
{\setbox0=\hbox{$\displaystyle\rm C$}\hbox{\hbox to0pt
{\kern0.4\wd0\vrule height0.9\ht0\hss}\box0}}
{\setbox0=\hbox{$\textstyle\rm C$}\hbox{\hbox to0pt
{\kern0.4\wd0\vrule height0.9\ht0\hss}\box0}}
{\setbox0=\hbox{$\scriptstyle\rm C$}\hbox{\hbox to0pt
{\kern0.4\wd0\vrule height0.9\ht0\hss}\box0}}
{\setbox0=\hbox{$\scriptscriptstyle\rm C$}\hbox{\hbox to0pt
{\kern0.4\wd0\vrule height0.9\ht0\hss}\box0}}}}
\def\Rl{{\mathchoice
{\setbox0=\hbox{$\displaystyle\rm R$}\hbox{\hbox to0pt
{\kern0.4\wd0\vrule height0.9\ht0\hss}\box0}}
{\setbox0=\hbox{$\textstyle\rm R$}\hbox{\hbox to0pt
{\kern0.4\wd0\vrule height0.9\ht0\hss}\box0}}
{\setbox0=\hbox{$\scriptstyle\rm R$}\hbox{\hbox to0pt
{\kern0.4\wd0\vrule height0.9\ht0\hss}\box0}}
{\setbox0=\hbox{$\scriptscriptstyle\rm R$}\hbox{\hbox to0pt
{\kern0.4\wd0\vrule height0.9\ht0\hss}\box0}}}}

\title{Anomaly-free formulation of non-perturbative, four-dimensional
Lorentzian quantum gravity}
\author{T. Thiemann\thanks{thiemann@math.harvard.edu} \\
       Physics Department, Harvard University, \\
       Cambridge, MA 02138, USA}
\date{{\small Preprint HUTMP-96/B-350}}

\begin{document}

\maketitle

\begin{abstract}
A Wheeler-Dewitt quantum constraint operator for four-dimensional, 
non-perturbative Lorentzian vacuum quantum gravity is defined in the 
continuum. The regulated Wheeler-DeWitt constraint operator is densely
defined, does not require any renormalization and the final operator is 
anomaly-free and at least symmmetric.\\
The technique introduced here can also be used to produce a couple of 
completely well-defined 
regulated operators including but not exhausting a) the Euclidean 
Wheeler-DeWitt operator, b)
the generator of the Wick rotation transform that maps solutions to 
the Euclidean Hamiltonian constraint to solutions to the Lorentzian 
Hamiltonian constraint, c) length operators, d) Hamiltonian
operators of the matter sector and e) the generators of the asymptotic 
Poincar\'e group including the quantum ADM energy. 
\end{abstract}

 
Attempts at defining an operator which corresponds to the Hamiltonian 
constraint of four-dimensional Lorentzian vacuum canonical gravity \cite{15} 
have first been made
within the framework of the ADM or metric variables (see, for 
instance, \cite{2}) This formulation of the theory seemed hopelessly difficult 
because of the complicated 
algebraic nature of the Hamiltonian (or Wheeler-DeWitt) constraint.
It was therefore thought to be mandatory to first cast the 
Hamiltonian constraint into polynomial form by finding better suited 
canonical variables. 
That this is indeed possible was demonstrated by Ashtekar \cite{7}.\\
There are two, a priori, problems with these Ashtekar connection 
variables for Lorentzian gravity : 1) they are complex valued and are 
therefore subject to algebraically
highly complicated reality conditions, difficult to impose on the quantum
level, which make sure that we are still 
dealing with real general relativity and 2) the Hamiltonian constraint
is polynomial only after rescaling it by a non-polynomial function, namely
a power of the square root of the determinant of the three-dimensional 
metric,  
that is, the original Wheeler-DeWitt constraint has actually been altered.\\
A solution to problem 1) has been suggested in \cite{3} (see also
\cite{10}) : namely,
one can define real Ashtekar variables \cite{8} which simplify the rescaled 
Hamiltonian constraint of Euclidean gravity and then construct a Wick rotation
transform from the Euclidean to the Lorentzian regime where the complex
Ashtekar variables are needed to simplify the Lorentzian constraint. The
advantage of these real-valued variables is that they allow for the 
construction of a mathematically rigorous kinematical framework by means
of which constraint operators can be regulated in a non-ambiguous fashion.
In particular, this framework has been successfully employed to arrive 
at the
complete solution of the Gauss and diffeomorphism constraints \cite{9}.
The main problem left is then to rigorously and consistently define the
Hamiltonian constraint. Given the developments in connection with the 
above mentioned Wick transform, in order 
to define then the Lorentzian Wheeler-DeWitt constraint operator it would 
be sufficient to define the Euclidean operator and the generator of the
transform.\\
However, concerning problem 2), apart from the non-appealing fact that the 
Wheeler-DeWitt constraint
was altered, what is worse is that the new Hamiltonian constraint carries 
density weight two. On general grounds, such an operator would need to be
renormalized (thus introducing a length scale) which breaks diffeomorphism 
invariance. A  solution to this
problem was first suggested in \cite{11} : the idea is to take the square
root of the rescaled Euclidean Hamiltonian constraint. While this seems to be
a required technical step to do in order to preserve diffeomorphism 
invariance there remain problems with it that have to do with taking
the square root of an infinite number of non-commuting, non-positive, not 
self-adjoint 
operators.\\ In this letter we introduce a novel technique which shows 
that \\
$\bullet$ The requirement that the Hamiltonian constraint should be
polynomial can be discarded. One {\em can} quantize the original
Wheeler-DeWitt constraint in its non-polynomial, unrescaled form in a 
satisfactory way. The resulting operator is surprisingly simple and the 
problem of computing its kernel is conceivable.  \\
$\bullet$ Since the Wheeler-DeWitt constraint carries density weight
one, problem 2) disappears, we never have to take ill-defined square roots.\\
$\bullet$ We never introduce complex variables and we never need to deal
with Euclidean gravity. A Wick rotation transform is completely
unnecessary : problem 1) also disappears. Quite surprisingly, however, the 
{\em unrescaled} Euclidean Hamiltonian constraint operator and the generator 
of the Wick transform appear very naturally in our regularization procedure 
so that we cannot avoid to construct them simultanously as a side result. 
This could turn out to be important for the task of actually solving the 
Wheeler-DeWitt constraint.\\
$\bullet$ It should be stressed at this point that all versions of 
Hamiltonian constraint operators for canonical quantum gravity in the 
continuum that have been constructed so far in the literature, whether
in the connection representation \cite{11a} or in the loop representation
\cite{11,13}, suffer from the fact that they provide quantizations only of
the Euclidean theory. Their validity relies heavily on the assumption that
one can actually exponentiate the generator of the Wick transform as 
outlined in \cite{3}. The mathematical difficulties associated with this
step are highly non-trivial and it is even possible that the Wick transform
simply does not exist. It is therefore the more important that we are able
to work directly with the Lorentzian constraint for which the correct 
reality conditions are implemented right from the beginning.\\
$\bullet$ The operator we obtain is not to be confused 
with the operator defined in \cite{30} (which is also formulated in terms of
real variables) because a) we work in the continuum rather than on a 
lattice and b) our operator is completely well-defined on the whole
Hilbert space while the one in \cite{30} suffers from singularities on a 
large subspace of it.\\

Let us fix the notation. 
Let the triad on the spacelike hypersurface $\Sigma$ be denoted by $e_a^i$,
where $a,b,c,...$ are tensorial and $i,j,k,...$ are $SU(2)$ indices.
The relation with the intrinsic metric is given by $q_{ab}=e_a^i e_b^j
\delta_{ij}$. It follows that $\det(q):=\det((q_{ab}))=[\det((e_a^i))]^2\ge 
0$. The
densitized triad is then defined by $E^a_i:=\det((e_b^j))e^a_i$ where
$e^a_i$ is the inverse of $e_a^i$. We also need the field $K_a^i=e K_{ab} 
e^b_i,\; e=\mbox{sgn}(\det((e_b^j))),$ where $K_{ab}$ is the extrinsic 
curvature of $\Sigma$. It turns
out that the pair $(K_a^i,E^a_i)$ is a canonical one, that is, these
variables obey
canonical brackets $\{K_a^i(x),E^b_j(y)\}=\delta^{(3)}(x,y)\delta_a^b
\delta_j^i$.\\
Let the spin-connection (which annihilates the triad) be denoted by
$\Gamma_a^i$. Then one can show that $(A_a^i:=\Gamma_a^i+K_a^i,E^a_i)$
is a canonical pair
on the phase space of Lorentzian gravity subject
to the $SU(2)$ Gauss constraint, the diffeomorphism constraint and 
the Wheeler-DeWitt constraint (neglecting a term proportional to the
Gauss constraint)
\be \label{1}
H:=\sqrt{\det(q)}[K_{ab}K^{ab}-\frac{1}{2}(K_a^a)^2-R]
=\frac{1}{\sqrt{\det(q)}}\mbox{tr}((F_{ab}-2R_{ab})[E^a,E^b])
\ee
where $F_{ab}$ and $R_{ab}$ respectively are the curvatures of the
$SU(2)$ connection $A_a^i$ and the triad $e_a^i$ respectively.\\
What has been gained by reformulating canonical gravity as a dynamical
theory of $SU(2)$ connections is the following : if, as we do in the sequel, 
one makes the  assumption that there exists a phase for quantum gravity in 
which the
excitations of the gravitational fields can be probed by loops rather than,
say, test functions of rapid decrease, then one has access to 
a powerful calculus on the space of (generalized) connections modulo
gauge transformations $\agb$ and, in particular, there is a natural
choice of a diffeomorphism invariant, faithful measure $\mu_0$ thereon which 
equips us with a Hilbert space ${\cal H}:=L_2(\agb,d\mu_0)$, appropriate 
for a representation in which $A$ is diagonal. Moreover, Gauss and 
diffeomorphism
constraints can be solved (see \cite{9} and references therein for
an introduction into these concepts).\\
The remaining step then is to give a rigorously defined quantum operator
corresponding to the Wheeler-DeWitt constraint and to project the scalar
product on its kernel. We do this in a series of three steps.

Step A) We begin by giving meaning to an operator corresponding
to the Euclidean Hamiltonian constraint
\be \label{2}
H^E:=\frac{1}{\sqrt{\det(q)}}\mbox{tr}(F_{ab}[E^a,E^b]) \;.
\ee
The method applied in \cite{11,11a} is to absorb the prefactor 
$1/\sqrt{\det(q)}$
into the lapse function and to give meaning to the operator corresponding to
the square root of the trace. 
There is also an approach \cite{13} that avoids taking the square root, 
however, then one discovers a singularity which needs to be renormalized and
this breaks diffeomorphism invariance. It can be recovered upon removing 
the regulator but then, to the best of our knowledge, all manipulations only 
have a quite formal character. Finally, in \cite{30} the Wheeler-DeWitt
constraint is multiplied by a power of $\sqrt{\det(q)}$ to render it 
polynomial but such a procedure only works on the lattice where the density
weight does not matter. In a final continuum limit one will ultimately
encounter singularities of even worse character than in \cite{13}.\\
By employing the method described below we can avoid the complications that 
arise in all of these approaches. The 
presentation will be brief, details will appear elsewhere \cite{1}.\\
We wish to impose the (self-adjoint) constraint operator on diffeomorphism 
invariant distributions $\psi$ on $\agb$, that is, we evaluate $\psi$ 
on the action of 
the constraint on so-called ``cylindrical" gauge invariant functions $f$
\cite{9} and require the resulting number to vanish, that is
$\psi(\hat{H}f)=0$ where $\psi(f)=\int_\agb d\mu_0 \overline{\psi}f$.
In brief terms, gauge invariant cylindrical functions on the space of 
(generalized) $SU(2)$ connections
are just finite linear combinations of traces of the holonomy around closed
loops in $\Sigma$. Each such function therefore may equally well be labeled
by the closed graph $\gamma$ consisting of the union of all loops involved
in that linear combination. 
Such a graph consists of a finite number of edges $e_1,..,e_n$ and vertices 
$v_1,..,v_m$. So, a function cylindrical with respect to a graph $\gamma$
typically looks like
$f(A)=f_\gamma(h_{e_1}(A),..,h_{e_n}(A))$, where $h_e(A)$ is the holonomy
along $e$ for the connection $A$ and $f_\gamma$ is a 
gauge invariant function on $SU(2)^n$.\\
We are now ready to explain the main idea.\\
$\bullet$ The total volume of $\Sigma$ is given by
\be \label{5}
V:=\int d^3x \sqrt{|\det(q)|} \;.
\ee
Since we will be using only the variation of $V$ it is understood 
that if $\Sigma$ is not compact then we first take a one parameter
family of bounded subsets $\Sigma_R\subset\Sigma=\Sigma_\infty$ 
where $\Sigma_R\subset\Sigma_{R'}$ for $R<R'$ 
to compute the variation of $\int_{\Sigma_R}\sqrt{|\det(q)|}$ and then
take the limit $R\to\infty$.\\
The first fact that we are going to use is that there is a well-defined,
self-adjoint operator $\hat{V}$ on $\cal H$ corresponding to $V$ \cite{5} 
whose action on 
cylindrical functions is perfectly finite : (following the second 
reference in \cite{5}) \be \label{6}
\hat{V}f=\left( \sum_{v\in V(\gamma)} 
\sqrt{|\frac{i}{8\cdot 3!}\sum_{e_I\cap e_J\cap e_K=v}
\epsilon(e_I,e_J,e_K)\epsilon_{ijk}
X^i_I X^j_J X^k_K|}\; \right) \;f_\gamma(g_1,..,g_n)
\ee
where $\epsilon(e_I,e_J,e_K)=
\mbox{sgn}(\det(\dot{e}_I(0),\dot{e}_J(0),\dot{e}_K(0)))$ and $V(\gamma)$
is the set of vertices of $\gamma$.
We have abbreviated $g_I=h_{e_I}(A)$ and $X_I=X(g_I)$ is the
right invariant vector field on $SU(2)$ (we have chosen orientations such
that all edges are outgoing at $v$). This demonstrates that $\hat{V}$
is a finite and well-defined operator on cylindrical functions.\\
$\bullet$ Secondly, we exploit the elementary chain of identities 
($\epsilon^{abc}$ has density weight one) 
\be \label{7}
\frac{[E^a,E^b]_i}{\sqrt{\det(q)}}
=\epsilon^{abc}e e_c^i(x)
=2\epsilon^{abc}\frac{\delta V}{\delta E^c_i(x)}
=2\epsilon^{abc}\{A_c^i,V\} \;. 
\ee
Then the Euclidean Hamiltonian constraint functional can be written ($N$ 
is the lapse function)
\be \label{9}
H^E[N]=2\int_\Sigma d^3x N(x)\epsilon^{abc}\mbox{tr}(F_{ab}\{A_c,V\}) \;.
\ee
We now triangulate $\Sigma$ into elementary tetrahedra $\Delta$ and for each
$\Delta$ we pick one of its vertices and call it $v(\Delta)$. Let 
$e_i(\Delta),
\;i=1,2,3$ be the three edges of $\Delta$ meeting at $v(\Delta)$. Let 
$\alpha_{ij}(\Delta):=e_i(\Delta)\circ a_{ij}(\Delta)\circ 
e_j(\Delta)^{-1}$ be the loop based at $v(\Delta)$ where $a_{ij}$ 
is the obvious other edge of $\Delta$ connecting 
those endpoints of $e_i,e_j$ which are distinct from $v(\Delta)$.
Then it is easy to see that 
\be \label{10}
H^E_\Delta[N]:=-\frac{2}{3}N_v\epsilon^{ijk}\mbox{tr}(h_{\alpha_{ij}(\Delta)}
h_{e_k(\Delta)}\{h_{e_k(\Delta)}^{-1},V\})
\ee
tends to $2\int_\Delta N\mbox{tr}(F\wedge\{A,V\})$ as we shrink 
$\Delta$
to the point $v(\Delta),\; N_v:=N(v(\Delta))$. Moreover, $H^E_\Delta[N]$ is 
gauge-invariant.\\ Let the triangulation be denoted by $T$. Then 
\be \label{11a}
H^E_T[N]=\sum_{\Delta\in T} H^E_\Delta[N]
\ee
is an expression which has the correct limit (\ref{9}) as all $\Delta$
shrink to their basepoints (of course the number of tetrahedra
filling $\Sigma$ grows under this process).\\
The reason for doing this is clear : If we now simply replace $V$
by $\hat{V}$ and the Poisson bracket by $1/i\hbar$ times the commutator
then 
\be \label{11}
\hat{H}^E_T[N]:=\sum_{\Delta\in T} \hat{H}^E_\Delta[N],\;
\hat{H}^E_\Delta[N]:=-\frac{2N_v}{3i\hbar}
\epsilon^{ijk}\mbox{tr}(h_{\alpha_{ij}(\Delta)}
h_{e_k(\Delta)}[h_{e_k(\Delta)}^{-1},\hat{V}])
\ee
is a regularized operator with the correct classical limit and whose 
action on cylindrical functions 
{\em is indeed finite} ! Namely, as may be suspected from the expression 
(\ref{6}), we find \cite{1}
\be  \label{12}
\hat{H}^E_T[N]f=\sum_{\Delta\in T;\; \Delta\cap\gamma\not=\emptyset}
\hat{H}^E_\Delta[N] f_\gamma, 
\ee
that is, a tetrahedron contributes to the action on $f$  only if it 
intersects the graph. Moreover, as one can show in more detail,
it only contributes if it intersects the graph in one of its vertices.
Therefore, if we choose our triangulations $T$ such that for each point
of $\Sigma$, the number of edges of tetrahedra of $T$ intersecting it,
is uniformly bounded (uniformly in $T$) from above by some integer 
then, no matter how fine the triangulations 
is, there are always only a finite number of terms involved in the sum 
(\ref{12}).\\
It is amazing that one got expression (\ref{11}) almost for free once
one knows that the volume operator is well-defined on holonomies,
no ill-defined products of distributions arise, we do not encounter
any singularities, no renormalization of the operator is 
necessary.\\ One can determine, for each graph, a triangulation which is 
diffeomorphism covariant \cite{1}, meaning that the prescription of how to 
attach the loops $\alpha_{ij}(\Delta)$ moves with the graph under 
diffeomorphisms in an appropriate sense.
Then, for reasons first observed in \cite{11}, when one evaluates a 
diffeomorphism invariant distribution $\psi$ on (\ref{12}), the number one 
gets depends only on the diffeomorphism class of the loop assignment 
$\alpha_{ij}(\Delta)$. Therefore, in this diffeomorphism invariant
context, the loops $\alpha_{ij}(\Delta)$ can be chosen as ``small" and the
triangulation as ``fine" as we wish, the value of (\ref{12}) on 
$\psi$ remains 
invariant and in that sense the continuum limit has already been taken.\\
However, the operator (\ref{11}) still carries a sign of the regularization
procedure : it depends on the diffeomorphism class $[T]$ of the 
triangulation assignment which labels the freedom that we have
in our regularization scheme.
It is therefore not an entirely trivial task to check whether
our operator $\hat{H}^E_T[N]$ is anomaly-free,
meaning that $[\hat{H}^E_T[M],\hat{H}^E_T[N]]f$ vanishes 
for any cylindrical function $f$ and lapse functions $M,N$ when evaluated
on a diffeomorphism invariant state. This severely constrains the freedom 
that we have in our choice of $[T]$ . Specifically,
a solution to the anomaly freedom problem is obtained {\em if all the loops
$\alpha_{ij}(\Delta)$ are chosen to be kinks with vertex at $v(\Delta)$} ! 
That is, the arc $a_{ij}(\Delta)$ joins the endpoints of 
$e_i(\Delta),e_j(\Delta)$ in an at least $C^1$ fashion. An arbitrary 
attachment of
$a_{ij}(\Delta)$ is insufficient to guarantee anomaly-freeness. 
In order to prove that the commutator vanishes we need to use 
diffeomorphism invariance as follows : notice that the action of the 
Euclidean Hamiltonian constraint is actually defined only up to a 
diffeomorphism. We need to make sure that for each choice of loop assignment
within the same diffeomorphism class for either of the operators
$\hat{H}^E(M),\hat{H}^E(N)$ and their products the commutator vanishes when
evaluated on a diffeomorphism invariant state $\psi$. When one performs
the calculation it turns out that for each such choices one gets a sum
of expressions of the form $k(M,N)[\hat{U}(\phi)f'-\hat{U}(\phi')f']$ where
$k(M,N)$ is a certain function depending on the lapses $M,N$ only, $f'$
is a function cylindrical with respect to a graph which is bigger than the
one that $f$ depended on, $\phi,\phi'$ are certain diffeomorphisms depending
on our choices of loop assignment and finally $\hat{U}(\phi)f_\gamma
=f_{\phi(\gamma)}$ is a unitary representation of the diffeomorphism group
$\mbox{Diff}(\Sigma)$ on $\cal H$. It is now obvious that $\psi$ vanishes on 
each of these expressions separately.\\ 
Finally, upon taking a symmetrical ordering of (\ref{11}) we manage to arrive
at a symmetric operator\footnote{actually a little more 
care is needed here \cite{1} : we need to make sure that the family of 
operators given
by (\ref{12}) is consistent \cite{20}, that is, they all are projections
to cylindrical subspaces of $\cal H$ of one and the same operator on 
$\cal H$}. Note that we have no problems 
in ordering $\hat{V}$ to the left or to the right of the holonomies 
involved since $\hat{V}$ has a finite action on holonomies of $A$ as is 
clear from (\ref{6}). The constraint algebra remains non-anomalous 
even after symmetric ordering which seems to be in conflict with
general arguments raised in \cite{Kuchar} for finite-dimensional models 
and which show that in symmetric
ordering the constraint algebra does never close with the generator of 
the diffeomorphism group appearing to the right of the structure functions. 
The resolution of the apparent contradiction is related to the fact that
the infinitesimal generator of $\hat{U}(\phi_t)$, where $\phi_t$ is a 
one parameter subgroup of $\mbox{Diff}(\Sigma)$, cannot be defined since
$\hat{U}(\phi_t)$ does not act strongly continuously on $\cal H$ \cite{9}. 
Therefore
the question of whether the diffeomorphism constraint operator appears  
to the right in the expression of the commutator of two Hamiltonian
constraints cannot even be asked.

Step B) Recall that the integrated (densitized) trace of the extrinsic 
curvature is up to a constant factor just the time derivative of 
the total volume with respect to the integrated Hamiltonian 
constraint (which is a signature invariant statement) : 
\be \label{13}
K:=\int_\Sigma d^3x \sqrt{\det(q)}K_{ab} q^{ab}=
-\int_\Sigma d^3x K_a^i E^a_i=-\{V,H^E[N=1]\} \ee
which also can be readily verified. Then we are naturally
led to define
\be \label{14}
\hat{K}_T:=\frac{-1}{i\hbar}[\hat{V},\hat{H}^E_T[1]] \;.
\ee
If we order $\hat{H}^E_T$ appropriately then both operators on the right 
hand side are symmetric and 
finite and we have produced a finite and symmetric expression for 
$\hat{K}_T$.

Step C) 
The aim is to write down an operator version of the
Wheeler-DeWitt constraint (\ref{1}) which is perfectly well-defined and 
finite. We have two strategies at our disposal.

Strategy 1)\\
Upon realizing the following identity 
\be \label{18}
K_a^i=\frac{\delta K}{\delta E^a_i}=\{A_a^i,K\},
\ee
the developments around step B) motivate to get rid of the complicated 
curvature term $R_{ab}$ involved in (\ref{1}) in favour of $K$. We have
\ba \label{19}
&& H+H^E=\frac{2}{\sqrt{\det(q)}}\mbox{tr}([K_a,K_b][E^a,E^b]) 
=\frac{2}{\sqrt{\det(q)}}\mbox{tr}([\{A_a,K\},\{A_b,K\}][E^a,E^b])
\nonumber\\
&& =4\epsilon^{abc}\mbox{tr}([\{A_a,K\},\{A_b,K\}]\{A_c,V\})
=8\epsilon^{abc}\mbox{tr}(\{A_a,K\}\{A_b,K\}\{A_c,V\})\;.
\ea
The last identity suggests to define the regularized Wheeler-DeWitt
operator in complete analogy with the Euclidean Hamiltonian 
constraint operator in the following, manifestly gauge invariant way
\be \label{20}
\hat{H}_T[N]:=-\left[\hat{H}^E_T[N]+\frac{8}{(i\hbar)^3}
\sum_{\Delta\in T}N_v\epsilon^{ijk}
\mbox{tr}(h_{e_i(\Delta)}[h_{e_i(\Delta)}^{-1},\hat{K}_T]
h_{e_j(\Delta)}[h_{e_j(\Delta)}^{-1},\hat{K}_T]
h_{e_k(\Delta)}[h_{e_k(\Delta)}^{-1},\hat{V}])\right]
\ee
and we see that the Wheeler-DeWitt
constraint operator can be built alone from the volume operator $\hat{V}$
and operators corresponding to holonomies along the edges of the 
tetrahedra of a triangulation.\\
As it stands it is not self-adjoint yet but it is clear that a 
symmetrical ordering can be performed (without picking up singularities) to 
render it symmetric and we expect it to possess self-adjoint extensions. \\
Finally, for the same reason that $\hat{H}^E_T[N]$ is anomaly-free, 
$\hat{H}_T$ is anomaly free as well.

Strategy 2)\\
This strategy simplifies the problem of finding
solutions to the constraints as (\ref{11}) is less complicated than 
(\ref{20}). It is a luxury at 
our disposal which we may use or not, however, it is not a necessary 
step.\\
The generator of the Wick rotation
transform can be defined now as $\hat{C}_T:=(\pi/2)\hat{K}_T$ and as argued
in \cite{3} we may just {\em define}
\be \label{15}
\hat{H}_T:=\hat{W}_T^{-1}\hat{H}^E_T\hat{W}_T,\mbox{ where } 
\hat{W}_T:=\exp(-1/\hbar \hat{C}_T)\;. \ee 
We would proceed by first finding solutions to $\hat{H}^E_T\psi_E=0$ 
and then just analytically continue them
to find solutions $\psi_\Co$ to the Lorentzian Wheeler-DeWitt constraint 
in a holomorphic representation. Such solutions are mapped unitarily \cite{3}
to solutions $\psi:=\hat{U}_T^{-1}\psi_\Co
=(\hat{W}_T)^{-1}\psi_E$ to $\hat{H}_T\psi=0$ in the real representation.
Expression (\ref{15}) has the disadvantage that 
whenever $\hat{H}^E_T$ is symmetric, $\hat{H}_T$ is not. The motivation for
having $\hat{H}_T$ symmetric is because we wish to find its kernel in the
form of generalized eigenvectors \cite{9}.\\
Expressions (\ref{20}), (\ref{15}) are anomaly-free, densely defined
operators corresponding to the original Wheeler-DeWitt operator in the 
continuum for non-perturbative four-dimensional Lorentzian canonical quantum 
vacuum gravity.

Some final comments are in order :\\
$\bullet$ Not even our Euclidean operator (\ref{11}) and the ones
proposed in \cite{13} and \cite{11,11a} have anything to do with each 
other, they are {\em entirely different}. The
only thing they share is that their classical limits or the square 
thereof are proportional to each other. It is therefore to be expected 
that the Euclidean solutions that have been found already in the literature 
for the operators defined in \cite{13,11} are far from being annihilated by our 
operator. What speaks for our operator is that a) none of the operators in
\cite{11,11a,13} can be used to define $\hat{K}$ along the lines 
proposed here because it was crucial that the classical identity 
$K=-\{V,H^E[1]\}$ holds and b) our procedure leads to the quantization of
the original Wheeler-DeWitt constraint, rather than a modified version
thereof.\\
$\bullet$ There is a lot of freedom involved in the regularization 
step reflecting the fact that the quantum theory of a given classical field
theory is not unique. An important, unresolved question is how to select the 
correct (or physically relevant) regularization of $\hat{H}$.\\
$\bullet$ As the Hamiltonian constraint operator on a given graph $\gamma$ 
reduces to
a finite number of {\em mutually commuting} (in the diffeomorphism 
invariant context), symmetric constraint 
operators (which we expect to have self-adjoint extensions), one for each 
vertex of 
$\gamma$, we can exponentiate it and it seems feasable to determine the
space of diffeomorphism invariant solutions to the Wheeler-DeWitt constraint
as well as a physical inner product thereon by the
group averaging method \cite{14,9}. To illustrate this, let the function 
$f_\gamma$
be cylindrical with respect to a graph $\gamma$ and denote by $V(\gamma)$
its set of vertices. Then the constraint equation has the 
structure 
$\hat{H}_T[N]f_\gamma=\sum_{v\in V(\gamma)}N(v)\hat{H}_{T,v}f_\gamma=0$ and
is formally solved by ($\mu_v$ is the Haar measure on the Abelian 
group generated by $\hat{H}_{T,v}$)
\be \label{24}
[f]_\gamma:=\prod_{v\in V(\gamma)} \int d\mu_v(a_v) e^{i a_v 
\hat{H}_{T,v}} \; f_\gamma 
\ee
Next, we take the group average \cite{9} over the graphs on which the 
decomposition of (\ref{24}) into cylindrical functions depends to obtain a 
solution $[f]_{[\gamma]}$
to both, the diffeomorphism and the Hamiltonian constraint and the 
physical inner product would be
\be \label{25}
<[f]_{[\gamma]},[f']_{[\gamma']}>_{phys}:=<[f]_{[\gamma]},f'_{\gamma'}>
\ee
where the second inner product is the one on $\cal H$. The construction
of interesting observables would parallel related procedures displayed 
in \cite{9}.\\
$\bullet$ The final expression of the Wheeler-DeWitt constraint (\ref{20})
is surprisingly simple : on each cylindrical function it is a low order
polynomial in the volume operator and holonomy operators and 
therefore one can hope to find exact solutions. 
While no solution could be found until now in closed 
form (except for cylindrical functions on two-valent graphs which, however, 
do not take the presence of the curvature term $F_{ab}$ fully into account)
the intuitive picture that arises concerning the action of the 
Wheeler-DeWitt constraint, is as follows : Recall \cite{12} that a 
spin-network state is a gauge invariant function $T_{\gamma,\vec{j}}(A)$ 
cylindrical with respect to a graph $\gamma$ where the dependence on its 
edges $e_1,..,e_n$ is through the matrix elements of irreducible 
representations of $SU(2)$ labeled by spins $\vec{j}=j_1,..,j_n$, evaluated 
at the holonomy along the corresponding edges. On such spin-network states 
the Hamiltonian constraint acts by 
annihilating, creating and re-routing the quanta of angular momentum 
associated with the edges of the graph in units of $\pm\hbar,\pm \hbar/2,0$. 
Remarkably,
the spectrum of the Hamiltonian constraint operator at a given vertex 
is largely determined by the spectrum of the volume operator so that it
becomes of utmost importance to gain control over it \cite{40}.\\
$\bullet$ It is clear that the method proposed here opens access
to other well-defined and finite operators which were so far out of reach
in a representation in which the intrinsic metric is not diagonal because we 
are able to make sense out of an operator corresponding to $q_{ab}$.
Examples are operators  
corresponding to the length of a curve \cite{4}, matter Hamiltonians
for canonical Yang-Mills theory \cite{25} or the ADM Hamiltonian
\cite{26}. It is extremely interesting to see whether the latter
Hamiltonian is at least positive semi-definite.\\
$\bullet$ We define the {\em spin-network representation} to be the abstract
representation defined by $<A|\gamma,\vec{j}>:=T_{\gamma,\vec{j}}(A)$
where as usual $<A'|A>=\delta_{\mu_0}(A',A)$. Then it can be shown that 
the ADM Hamiltonian acts effectively by multiplying the states 
$|\gamma,\vec{j}>$ by certain algebraic factors depending on the spins 
$\vec{j}$. In other words, the spin-network representation can be 
interpreted as the ``non-linear Fock-" (or occupation number) 
representation for quantum gravity. Roughly, the spins associated with 
the edges of a graph indicate the ``number" of elementary 
string-like excitations of the gravitational field along the edges of the 
graph very much like the integers indicate the number of photons 
associated with a certain array of excited modes for QED (see \cite{1}
for a more detailed discussion). 
\\
\\
{\large Acknowledgements}\\
\\
This research project was supported in part by DOE-Grant 
DE-FG02-94ER25228 to Harvard University.

\end{document}